# Production and Loss Processes of Hydrogen Energetic Neutral Atoms in the Heliosphere from 5 eV to 500 keV

Paweł Swaczyna*, Maciej Bzowski, and Marzena A. Kubiak

Space Research Centre PAS (CBK PAN), Bartycka 18a, 00-716 Warsaw, Poland

## Abstract

Energetic Neutral Atom (ENA) observations provide valuable insights into the plasma conditions in the heliosphere and the surrounding interstellar medium. Unlike plasma detectors, which measure charged particles tied to the magnetic fields at their location, ENA detectors capture former ions that were neutralized in distant regions and traverse the heliosphere in straight trajectories. ENA fluxes near the Sun represent line-of-sight integrals of parent ion fluxes multiplied by neutralization (production) rates and reduced by the probability of ENA reionization (loss) processes. So far, most ENA analyses have focused on charge exchange between hydrogen atoms and protons as the primary source of ENAs. Here, we examine various ENA production and loss processes throughout the heliosphere in the broad energy range (5 eV to 500 keV) encompassing the next-generation ENA instruments aboard the Interstellar Mapping and Acceleration Probe (IMAP) mission. Our study considers binary collisions involving the most abundant species: protons, electrons, α-particles, $He^+$ ions, photons, as well as hydrogen and helium atoms. Our findings indicate that, in addition to ENAs produced by charge exchange of energetic protons with hydrogen atoms, a significant portion of high-energy ENAs originate from the charge exchange with helium atoms. Below 10 keV, the dominant ENA loss processes are charge exchange collisions with protons and photoionization. However, stripping ionization processes, e.g., from collisions with ambient interstellar neutral hydrogen, become the main loss mechanism for higher energies because the charge exchange rate rapidly decreases.

## 1. Introduction

An energetic neutral atom (ENA) is formed when an energetic positive ion gains an electron, e.g., in a charge exchange collision with an ambient neutral atom. Contrary to charged particles, whose motions are defined by ambient electromagnetic fields, ENAs move almost freely through space, with the mean free path for reionization comparable to the size of the heliosphere. Because the gravity and radiation pressure affect their trajectories negligibly for energies above a few hundred eV (Bzowski 2008), ENAs travel long distances on almost straight trajectories through the heliosphere, enabling remote sensing of distant regions of the heliosphere by instruments near the Sun (Hsieh et al. 1992; Gruntman 1997; Hsieh & Möbius 2022). Therefore, while plasma instruments allow for studies of the distribution functions of particles only in the immediate surroundings of the spacecraft, ENA detectors provide an image of the global heliospheric structure (e.g., McComas et al. 2009a; Reisenfeld et al. 2021).

Each ENA is characterized by its energy and velocity direction, but the actual place of origin, i.e., the point in space where the parent ion was neutralized, may be located anywhere along the backtracked trajectory. Therefore, the ENA flux coming from a given direction consists of former ions neutralized along this trajectory, depleted underway by loss processes. This paper considers hydrogen ENAs created from energetic protons. The key production and loss process of hydrogen ENAs is the charge exchange between protons and hydrogen atoms. The main collision partners for the neutralization of energetic protons are ambient neutral atoms penetrating the heliosphere from the very local interstellar medium (VLISM). On the other hand, ENAs are lost due to collisions with solar wind protons. Nevertheless, it is not the only process neutralizing or ionizing the atoms, therefore, we perform a systematic discussion of possible production and loss sources.

* Corresponding author (pswaczyna@cbk.waw.pl)

The advent of ENA detectors able to measure ENA fluxes from the entire sky provided means to study the global structure of the heliosphere. The Interstellar Boundary Explorer (IBEX, McComas et al. 2009b) is the first mission dedicated to observations of heliospheric ENAs. IBEX is equipped with two ENA detectors covering the energy range from ~10 eV to 2 keV (IBEX-Lo, Fuselier et al. 2009) and from ~380 eV to 6 keV (IBEX-Hi, Funsten et al. 2009). The mission provided yearly maps of ENA emission for more than a full solar cycle (McComas et al. 2009a, 2012, 2014, 2017, 2020, 2024). Higher energy ENAs, from 5.2 to 55 keV, were measured by the Ion and Neutral Camera (INCA, Krimigis et al. 2004) on board the Cassini mission from 2000 to 2016 (Dialynas et al. 2013, 2019; Westlake et al. 2020). The Interstellar Mapping and Acceleration Probe (IMAP, McComas et al. 2018) is equipped with three ENA instruments covering an energy range spanning 5 orders of magnitude: IMAP-Lo (5 eV – 1 keV), IMAP-Hi (410 eV – 15.6 keV), and IMAP-Ultra (3 – 300 keV). The optimized design of this new generation of detectors provides better energy and angular resolution of ENA maps.

This study discusses the hydrogen ENA production and loss processes over the energy range covering all three IMAP ENA detectors, providing a comprehensive guide on the importance of the processes in the range of possible physical conditions in the heliosphere.

## 2. Definitions

The ENA differential flux $j_{\mathrm{ENA}}(E, \boldsymbol{\Omega})$, typically expressed in the unit of cm$^{-2}$s$^{-1}$sr$^{-1}$keV$^{-1}$, at energy $E$ seen from direction $\boldsymbol{\Omega}$ is given by the integral along a backtracked trajectory (e.g., Gruntman 1997; Zirnstein et al. 2013):

$$j_{\mathrm{ENA}}(\boldsymbol{r}_0, E, \boldsymbol{\Omega}) = \int_0^\infty j_{\mathrm{ion}}(\boldsymbol{r}_0 + r\boldsymbol{\Omega}, E, \boldsymbol{\Omega}) \left[\sum\nolimits_k \beta_k^{\mathrm{P}}(\boldsymbol{r}_0 + r\boldsymbol{\Omega}, E)\right] w(\boldsymbol{r}_0, E, \boldsymbol{\Omega}, r) \frac{dr}{v}, \tag{1}$$

where $v = \sqrt{2E/m}$ is the ENA speed, and $m$ is its mass. The formula relates the ENA differential flux $j_{\mathrm{ENA}}(\boldsymbol{r}_0, E, \boldsymbol{\Omega})$ at the detector located at $\boldsymbol{r}_0$ with the parent ion differential flux $j_{\mathrm{ion}}(\boldsymbol{r}_0 + r\boldsymbol{\Omega}, E, \boldsymbol{\Omega})$ along the line of sight. The ion flux in the integrand is in the spacecraft reference frame, which is calculated from the phase space distribution function $f_{\mathrm{ion}}(\boldsymbol{r}_0 + r\boldsymbol{\Omega}, -v\boldsymbol{\Omega})$:

$$j_{\mathrm{ion}}(\boldsymbol{r}_0 + r\boldsymbol{\Omega}, E, \boldsymbol{\Omega}) = \frac{v^2}{m} f_{\mathrm{ion}}(\boldsymbol{r}_0 + r\boldsymbol{\Omega}, -v\boldsymbol{\Omega}). \tag{2}$$

The second factor in the integrand is the sum of the collision rates $\beta_k^{\mathrm{P}}(\boldsymbol{r}_0 + r\boldsymbol{\Omega}, E)$ of all production processes, enumerated by $k$. The last factor is the survival probability of an ENA between the production location and the detector:

$$w(\boldsymbol{r}_0, E, \boldsymbol{\Omega}, r) = e^{-\int_0^r \sum_i \beta_i^{\mathrm{L}}(\boldsymbol{r}_0 + r'\boldsymbol{\Omega}, E) \frac{dr'}{v}} \tag{3}$$

We sum up the collision rates $\beta_i^{\mathrm{L}}(\boldsymbol{r}_0 + r'\boldsymbol{\Omega}, E)$ of all ionization processes enumerated by $i$. The collision rates are expressed as probabilities per unit of time, and in principle, the integration needs to be done over the time of flight to the detector. However, we transform the integration over time to integration over distances $r$ and $r'$ using the ENA velocity in the integrals in Equations (1) and (3).

Equation (1) assumes that ENAs move along straight lines. The ion and ENA fluxes in Equation (1) must be considered in the same reference frame. In practical applications, the ENA and ion fluxes are considered in the solar and plasma frames, respectively. Therefore, the appropriate transformations are needed. The ENA flux transformation from the spacecraft frame to the solar-inertial frame is called the Compton-Getting correction (McComas et al. 2010). The ion flux transformation to the plasma frame is generally more complicated as the plasma flow needs to be obtained from modeling (see Zirnstein et al. 2021 for detailed discussion).

The production and losses of ENAs are often considered to be spatially separated. The production of ENAs observed by IBEX or IMAP takes place mostly beyond the termination shock in the heliosheath, but the losses occur mostly inside the termination shock. If so, the lower boundary of the integral in Equation (1) and the upper boundary of the integral in Equation (3) are often approximated as the termination shock distance. In such a case, the survival probability can be factorized outside the main integral and calculated separately from the production integral. Nevertheless, for distant ENA sources, e.g., such as the IBEX ribbon, losses beyond the termination shock should be accounted for in the modeling.

An appropriate cross section is needed to calculate the collision rate for a given process. In classical physics, the cross section is an area around one of the interacting particles perpendicular to their relative motion, inside which the reaction occurs. Cross sections are typically denoted by σ and expressed in $cm^2$. For a test particle traversing a homogeneous medium where the phase space distribution of collision partners is given by $nf(\boldsymbol{v})$, the collision rate is:

$$\beta = \int \sigma(|\boldsymbol{v}_\mathrm{a} - \boldsymbol{v}|)\,|\boldsymbol{v}_\mathrm{a} - \boldsymbol{v}| n f(\boldsymbol{v}) d^3\boldsymbol{v}, \tag{4}$$

where $\boldsymbol{v}_\mathrm{a}$ is the velocity vector of the test particle. The density, $n$, can be factorized from the distribution function, and the collision rate normalized by the density is called the collision rate coefficient, typically denoted by $\alpha$ or $\langle \sigma v_\mathrm{rel} \rangle$:

$$\alpha \equiv \langle \sigma v_\mathrm{rel} \rangle \stackrel{\mathrm{def}}{=} \frac{\beta}{n}. \tag{5}$$

The symbol $\langle \sigma v_\mathrm{rel} \rangle$ is used to emphasize that the collision rate coefficient is the average of the product of the cross section and relative speed. Most generally, the integral in Equation (4) needs to be evaluated numerically because the cross section may have a complicated functional form. However, if the thermal speed of the collision partners is small compared to the difference between the test particle velocity and the bulk velocity of the collision partners, this integral can be approximated as:

$$\langle \sigma v_\mathrm{rel} \rangle \approx \sigma(v_\mathrm{rel}) v_\mathrm{rel}, \tag{6}$$

where $v_\mathrm{rel}$ is the mean relative speed. If the collision partners are distributed according to the Maxwell-Boltzmann distribution, this speed is given by (Pauls et al. 1995)[†]:

[†] This formula with a typo was also given in Ripken & Fahr (1983).

$$v_{\mathrm{rel}} = v_{\mathrm{th}} \left[\frac{1}{\sqrt{\pi}} \exp(-x^2) + \left(x + \frac{1}{2x}\right) \mathrm{erf}(x)\right] \approx v_{\mathrm{th}} \sqrt{\frac{4}{\pi} + x^2}, \tag{7}$$

where $v_{\mathrm{th}} = \sqrt{2kT/m}$ is the thermal speed, $x = |\boldsymbol{v}_a - \boldsymbol{u}|/v_{\mathrm{th}}$ is the normalized bulk speed, $\boldsymbol{u}$ – is the bulk speed of the collision partners, $k$ – the Boltzmann constant, $T$ – the temperature of the collision partners, and $m$ – the mass of the collision partner. The maximum error of the approximate formula on the right-hand side of Equation (5) is less than 2.5% close to $x = 1.3$, but the formula is accurate for both $x \gg 1$ and $x \ll 1$.

The approximation given in Equation (6) is frequently used in space plasma applications. It is accurate if the terms proportional to $(v - v_{\mathrm{rel}})^2$ and higher orders in the Taylor expansion of the cross section in the proximity of $v_{\mathrm{rel}}$ defined by the thermal speed $v_{\mathrm{th}}$ are negligible. DeStefano & Heerikhuisen (2020) derived more complex formulae for the case with non-negligible terms in the Taylor expansion and generalization for the kappa distributions (e.g., Livadiotis & McComas 2013).

Cross sections depend on the collision energy. The collision energy can be expressed as the energy of one of the reactants (projectile) in the reference frame of the other reactant (target) or as the center-of-mass energy, in which the energies of both particles in the center-of-mass reference frame are combined. As the energy expressed, e.g., in eV, depends on the definition, we normalize the projectile's energy by the projectile's mass and the center-of-mass energy by the reduced mass. The normalized energies are unambiguous and expressed in eV/u or keV/u, where u represents the unified atomic mass unit. Note that the atomic mass unit is close to 1 u for hydrogen atoms and protons.

The rate calculated from Equation (5) is the probability of a collision per unit time. Nevertheless, for the collision probability of a neutral particle, e.g., for the consideration of ENA losses, it is useful to express this probability as the exposure along the particle trajectory:

$$\epsilon_k = \int_{t_{\mathrm{b}}}^{t_{\mathrm{e}}} \beta_k dt = \int_{\boldsymbol{r}_{\mathrm{b}}}^{\boldsymbol{r}_{\mathrm{e}}} \frac{\beta_k}{|\boldsymbol{u}_a|} dr\,, \tag{8}$$

where the exposure $\epsilon_k$ is calculated in the period between $t_{\mathrm{b}}$ and $t_{\mathrm{e}}$. The right-hand part of the equation assumes a straight-line trajectory with the particle moving at a constant speed $|\boldsymbol{u}_a|$ between positions $\boldsymbol{r}_{\mathrm{b}}$ and $\boldsymbol{r}_{\mathrm{e}}$. This approximation is valid for most ENAs but for slower atoms, such as ISN atoms, the velocity changes near the Sun. The survival probability of an ENA traveling between the points $\boldsymbol{r}_{\mathrm{b}}$ and $\boldsymbol{r}_{\mathrm{e}}$ is calculated from the sum of exposures to relevant loss processes:

$$w = \exp(-\epsilon) = \exp\left(-\sum_k \epsilon_k\right). \tag{9}$$

For small exposures ($\epsilon \ll 1$), the first-order approximation of the survival probability is $w \approx 1 - \epsilon$. Therefore, small values of $\epsilon$ are equivalent to the ionization probability over the considered path. The exposure factors for different ionization processes should be summed together, or, equivalently, the survival probabilities of loss processes need to be multiplied with each other.

## 3. Representative Physical Conditions in the Heliosphere

To estimate the importance of ENA production and loss processes, we define six exemplary condition sets representing various regions of the heliosphere. Nevertheless, the actual calculations of the production and losses in interpreting the observations should employ comprehensive models providing properties of the plasma and neutral components. Table 1 summarizes the properties of these representative regions. For all regions, we assume quasi-neutrality of the plasma and neglect any species heavier than helium. Therefore, the electron number density is $n_{\mathrm{e}} = n_{\mathrm{p}} + n_{\mathrm{He}^+} + 2n_{\alpha} + n_{\mathrm{HPI}} + n_{\mathrm{HePI}}$ (see Table 1 for symbol definitions). We round most values to one or two significant digits owing to their uncertainty and variation over time. We neglect other species because their densities are low, and thus, they cannot contribute significantly to the production or loss of ENAs.

**Table 1. Representative Sets of Physical Conditions in the Heliosphere**

| | **SSW 1 au** | **FSW 1 au** | **SSW 45 au** | **FSW 45 au** | **Heliosheath** | **VLISM** |
|---|---|---|---|---|---|---|
| | | | *Charged particles* | | | |
| **Bulk speed ($v_{\mathrm{pl}}$)** | 400 km s$^{-1}$ | 750 km s$^{-1}$ | 400 km s$^{-1}$ | 750 km s$^{-1}$ | 100 km s$^{-1}$ | 15 km s$^{-1}$ |
| **$H^+$ density ($n_{\mathrm{p}}$)** | 6 cm$^{-3}$ | 1.7 cm$^{-3}$ | 3×10$^{-3}$ cm$^{-3}$ | 8×10$^{-4}$ cm$^{-3}$ | 4×10$^{-3}$ cm$^{-3}$ | 0.1 cm$^{-3}$ |
| **$H^+$ th. speed ($u_{\mathrm{p}}$)** | 30 km s$^{-1}$ | 30 km s$^{-1}$ | 13 km s$^{-1}$ | 13 km s$^{-1}$ | 34 km s$^{-1}$ | 18 km s$^{-1}$ |
| **$He^{2+}$ density ($n_{\alpha}$)** | 0.2 cm$^{-3}$ | 0.07 cm$^{-3}$ | 9×10$^{-5}$ cm$^{-3}$ | 3×10$^{-5}$ cm$^{-3}$ | 1.2×10$^{-4}$ cm$^{-3}$ | … |
| **$He^{2+}$ th. speed ($u_{\alpha}$)** | 15 km s$^{-1}$ | 15 km s$^{-1}$ | 6 km s$^{-1}$ | 6 km s$^{-1}$ | 17 km s$^{-1}$ | … |
| **$He^+$ density ($n_{\mathrm{He+}}$)** | … | … | 1.4×10$^{-6}$ cm$^{-3}$ | 2×10$^{-6}$ cm$^{-3}$ | 1.2×10$^{-5}$ cm$^{-3}$ | 0.017 cm$^{-3}$ |
| **$He^+$ th. speed ($u_{\mathrm{He+}}$)** | … | … | 6 km s$^{-1}$ | 6 km s$^{-1}$ | 17 km s$^{-1}$ | 9 km s$^{-1}$ |
| **$e^-$ density ($n_{\mathrm{e}}$)** | 6.0/0.4 cm$^{-3}$[(a)] | 1.7/0.1 cm$^{-3}$[(a)] | 4×10$^{-3}$ cm$^{-3}$ | 1×10$^{-3}$ cm$^{-3}$ | 6×10$^{-3}$ cm$^{-3}$ | 0.12 cm$^{-3}$ |
| **$e^-$ th. speed ($u_{\mathrm{e}}$)** | 1700/5500 km s$^{-1}$ | 1700/5500 km s$^{-1}$ | 550 km s$^{-1}$ | 550 km s$^{-1}$ | 1500 km s$^{-1}$ | 780 km s$^{-1}$ |
| **$H^+$ PUI density ($n_{\mathrm{HPI}}$)** | … | … | 4×10$^{-4}$ cm$^{-3}$ | 1×10$^{-4}$ cm$^{-3}$ | 1×10$^{-3}$ cm$^{-3}$ | … |
| **$H^+$ PUI th. speed ($u_{\mathrm{HPI}}$)** | … | … | 270 km s$^{-1}$ | 500 km s$^{-1}$ | 400 km s$^{-1}$ | … |
| **$He^+$ PUI density ($n_{\mathrm{HePI}}$)** | … | … | 1.2×10$^{-5}$ cm$^{-3}$ | 7×10$^{-6}$ cm$^{-3}$ | 3×10$^{-5}$ cm$^{-3}$ | … |
| **$He^+$ PUI th. speed ($u_{\mathrm{HePI}}$)** | … | … | 270 km s$^{-1}$ | 500 km s$^{-1}$ | 400 km s$^{-1}$ | … |
| | | | *Neutral atoms* | | | |
| **$H^0$ bulk speed ($v_{\mathrm{H}}$)** | 22 km s$^{-1}$ | 22 km s$^{-1}$ | 22 km s$^{-1}$ | 22 km s$^{-1}$ | 22 km s$^{-1}$ | 22 km s$^{-1}$ |
| **$H^0$ density ($n_{\mathrm{H}}$)** | 3×10$^{-3}$ cm$^{-3}$ | 3×10$^{-3}$ cm$^{-3}$ | 0.11 cm$^{-3}$ | 0.11 cm$^{-3}$ | 0.15 cm$^{-3}$ | 0.25 cm$^{-3}$ |
| **$H^0$ th. speed ($u_{\mathrm{H}}$)** | 14 km s$^{-1}$ | 14 km s$^{-1}$ | 14 km s$^{-1}$ | 14 km s$^{-1}$ | 14 km s$^{-1}$ | 14 km s$^{-1}$ |
| **$He^0$ bulk speed ($v_{\mathrm{He}}$)** | 50 km s$^{-1}$ | 50 km s$^{-1}$ | 25 km s$^{-1}$ | 25 km s$^{-1}$ | 25 km s$^{-1}$ | 25 km s$^{-1}$ |
| **$He^0$ density ($n_{\mathrm{He}}$)** | 0.01 cm$^{-3}$ | 0.01 cm$^{-3}$ | 0.015 cm$^{-3}$ | 0.015 cm$^{-3}$ | 0.015 cm$^{-3}$ | 0.015 cm$^{-3}$ |
| **$He^0$ th. speed ($u_{\mathrm{He}}$)** | 5 km s$^{-1}$ | 5 km s$^{-1}$ | 5 km s$^{-1}$ | 5 km s$^{-1}$ | 5 km s$^{-1}$ | 5 km s$^{-1}$ |

**Notes:** [a]two electron populations (core/halo) at 1 au.

The **slow solar wind at 1 au (SSW 1 au)** represents the conditions near Earth's orbit. The adopted solar wind speed of 400 km $s^{-1}$ and density of 6 $cm^{-3}$ are typical values for the entire solar cycle 24 based on observations collected in the OMNI collection[‡] (see, e.g., Sokół et al. 2020).

The relative abundance of α-particles depends on the solar cycle phase and solar wind speed and changes between 1% and 5% (Kasper et al. 2007). We adopt 3%, i.e., the typical value for solar minima. We neglect $He^+$ ions because they are observed at 1 au only in connection with coronal mass ejections (Pissarenko et al. 1985; Skoug et al. 1999). We assume the solar wind temperature of 50,000 K, corresponding to the thermal speeds of about 30 and 15 km $s^{-1}$ for protons and α-particles, respectively. The velocity distribution function of electrons at 1 au is non-Maxwellian. Following the model adopted by Bzowski (2008), we approximate this distribution with two Maxwell components: core (94%) and halo (6%) with temperatures of $10^5$ and $10^6$ K (thermal speeds of 1700 and 5500 km $s^{-1}$), respectively.

We also include ISN populations. The ISN hydrogen density at 1 au is significantly reduced compared to the density at the termination shock. During the solar minimum, the density is between 0.002 (downwind) and 0.02 (upwind) of the density at the termination shock (Ruciński et al. 1996; Sokół et al. 2019). Assuming the upwind filtration and the termination shock density of 0.127 $cm^{-3}$ (Swaczyna et al. 2020), the density at 1 au is 0.003 $cm^{-3}$. We assume that the ISN atoms are described by one population combining primary and secondary atoms. Based on SWAN/SOHO observations (Lallement et al. 2005), we adopt the flow speed of 22 km $s^{-1}$ and temperature of 11,500 K (thermal speed of 14 km $s^{-1}$). The ionization of ISN helium is smaller, but the density is increased in the focusing cone. We use the density of 0.01 $cm^{-3}$ based on ~33% reduction in the upwind direction during the solar minimum (Sokół et al. 2019) and the density far from the Sun of 0.015 $cm^{-3}$ (Gloeckler & Geiss 1998). We use the bulk velocity of 50 km $s^{-1}$ owing to the gravitational acceleration in the heliosphere (Swaczyna et al. 2023a) and temperature of 7500 K, corresponding to the thermal speed of 5 km $s^{-1}$ (Swaczyna et al. 2023b). We neglect the contribution from the secondary ISN helium population.

The **fast solar wind at 1 au (FSW 1 au)** represents the conditions at higher heliographic latitudes during solar minima. We adopt the solar wind speed of 750 km $s^{-1}$ based on the Sokół et al. (2020) model. Assuming that the solar wind dynamic pressure is approximately the same as for the slow solar wind described above (McComas et al. 2008), we select the proton density of 1.7 $cm^{-3}$. For this case, we assume a higher, 4%, α-particle contribution as typical for faster solar wind (Kasper et al. 2007) and higher heliographic latitudes (McComas et al. 2000). We adopt the same temperatures as in the slow solar wind. The neutral components are assumed to be identical to the slow solar wind case.

The **slow solar wind at 45 au (SSW 45 au)** represents the conditions in the supersonic solar wind approximately halfway to the termination shock. We assume that the plasma bulk velocity is unchanged from 1 au and that the density decreases as $r^{-2}$, thus the proton density is ~$3\times10^{-3}$ $cm^{-3}$. We assume the plasma temperature of $10^4$ K (corresponding to thermal speeds of 13 and 6 km $s^{-1}$ for protons and α-particles, respectively) based on the New Horizons/SWAP observations (McComas et al. 2021). We note that the SWAP observations indicate slightly higher density at 45 au (McComas et al. 2021), possibly due to transient phenomena.

[‡] https://omniweb.gsfc.nasa.gov/

Swaczyna et al. (2019a) showed that for solar speed of 400 km $s^{-1}$, approximately 1.5% of α-particles are partially neutralized, forming a population of $He^+$ ions. Assuming the same content of α-particles as for the 1 au case, we obtain the density of α-particles and $He^+$ ions of ~$9\times10^{-5}$ and ~$1.4\times10^{-6}$ $cm^{-3}$, respectively. Fraternale et al. (2023) argued that their temperature is likely comparable to that of the thermal protons in the distant supersonic solar wind. Therefore, the electron temperature is also $10^4$ K, and the thermal speed is ~550 km $s^{-1}$.

PUIs accumulate in the expanding solar wind, and their number density is ~12% of the solar wind protons (McComas et al. 2021). Therefore, we assume the PUI density of ~$4\times10^{-4}$ $cm^{-3}$. The PUI temperature is ~$4.4\times10^6$ K (McComas et al. 2021), corresponding to a thermal speed of ~270 km $s^{-1}$. Additionally, the ionization of ISN helium atoms produces $He^+$ PUIs. The main ionization process is the photoionization. With the photoionization rate at 1 au of $10^{-7}$ $s^{-1}$ and using the procedure described in Section 2.3.2 in Swaczyna et al. (2017), we estimate the $He^+$ PUI density as $1.2\times10^{-5}$ $cm^{-3}$ at 45 au.

The neutral components at 45 au are not yet significantly affected by the processes operating close to the Sun. Based on the SWAP observations, the ISN hydrogen density at 45 au is ~0.11 $cm^{-3}$ (Swaczyna et al. 2020, 2024). The ISN helium density is assumed to be equal to the density far from the Sun of 0.015 $cm^{-3}$ (Gloeckler & Geiss 1998). The ISN helium atoms are not yet significantly accelerated in the solar gravity field. Thus, their bulk speed is assumed to be 25 km $s^{-1}$.

The **fast solar wind at 45 au (FSW 45 au)** conditions are derived assuming a fast solar wind stream moving at 750 km $s^{-1}$. We use the same procedure to estimate the densities of the components. For higher solar wind speeds, the number of α-particles neutralized to $He^+$ ions is ~7% (Swaczyna et al. 2019a). Assuming that the relative abundance of $H^+$ PUIs is the same as in the slow solar wind, their density is ~$1\times10^{-4}$ $cm^{-3}$. We assume the same temperature for core solar wind components as in the slow solar wind, but the thermal speed of $H^+$ PUIs is assumed to scale proportionally to the bulk solar wind speed because they are created at a proportionally higher injection speed. The $He^+$ PUI density is estimated using the same method as in the slow solar wind case. The ISN properties are assumed to be the same as in the slow solar wind.

The **heliosheath** is the region of the heliosphere between the termination shock and the heliopause. The physical conditions in the heliosheath vary significantly. Nevertheless, we intend our parameter set to reflect the average conditions between the termination shock and the heliopause in the forward part of the heliosphere. We use Voyagers' observations and modeling in similar directions. Both Voyagers traversed the heliosheath in the upwind hemisphere at moderate heliographic latitudes. The plasma bulk velocity observed by Voyager 1 shortly after the termination shock crossing was ~150 km $s^{-1}$ and decreased to ~80 km $s^{-1}$ near the heliopause crossing (Richardson et al. 2009, 2019). While the plasma instrument is not operational on Voyager 2, observations of energetic ions can indirectly determine the plasma flow (Krimigis et al. 2019). While the accuracy of this method may be questioned, the general range of predicted radial speeds was from zero to ~100 km $s^{-1}$ (Richardson et al. 2021). We adopt the bulk plasma speed of 100 km $s^{-1}$ for our study.

The proton density and temperature observed by the plasma instrument on Voyager 1 in the heliosheath were between 0.002 and 0.006 $cm^{-3}$ and 50,000 and 90,000 K (Richardson et al. 2019), respectively. We use the mean values of these ranges, i.e., the density of 0.004 $cm^{-3}$ and temperature of 70,000 K. We adopt the same temperature for protons, helium ions, and electrons. Furthermore, we include 3% of solar wind helium ions, from which 90% are assumed to be α-particles and 10% are singly neutralized $He^+$ ions. The expected density ratio of $H^+$ PUIs to solar wind protons at the termination shock is ~25% (McComas et al.

2021). Assuming that the production of new PUIs is smaller in the heliosheath, we use this number to estimate the $H^+$ PUI density of 0.001 $cm^{-3}$. PUIs were not measured by Voyagers at the termination shock. Therefore, we use an estimate of the PUI temperature of $10^7$ K, corresponding to the thermal speed of ~400 km $s^{-1}$, based on modeling (Bera et al. 2023). Furthermore, we adopt the mean relative $He^+$ to $H^+$ PUI density ratio from 45 au to estimate the $He^+$ PUI density.

The neutral components are assumed with the same parameters as at 45 au, except for ISN hydrogen density that is assumed to be higher, ~0.15 $cm^{-3}$, to account for the density ramp in the heliosheath as predicted by models (Swaczyna et al. 2024).

The **VLISM** within a few hundred au beyond the heliopause is noticeably modified by the presence of the heliosphere. As with the heliosheath, modeling suggests a strong spatial variation of the physical conditions. Nevertheless, we want to represent the conditions within the hydrogen wall along the Voyagers' trajectories. The flow speed is not directly measured, but we assume it is 15 km $s^{-1}$, i.e., slower than the pristine VLISM speed. The electron density derived from detecting plasma oscillations on Voyager 1 suggests a density from ~0.06 to ~0.14 $cm^{-3}$. We use a value of 0.12 $cm^{-3}$ for our considerations. Assuming that the dominant ions in the VLISM are protons and $He^+$ ions and using the number density ratio of ~6:1 of these species (Bzowski et al. 2019), we adopt the densities of protons and $He^+$ of 0.1 $cm^{-3}$ and 0.017 $cm^{-3}$, respectively. We use the plasma temperature of 20,000 K based on global modeling (Fraternale et al. 2021), corresponding to the thermal speeds of 18, 9, and 780 km $s^{-1}$ for protons, $He^+$ ions, and electrons, respectively. Finally, we use a hydrogen density of 0.25 $cm^{-3}$ to represent higher density in the VLISM, while the other neutral population parameters remain unchanged.

## 4. Production of Hydrogen ENAs

Production of ENAs from protons requires collisions with particles with electrons. Among the considered species, collisions with the following particles may produce an ENA: $H^0$, $He^0$, $He^+$, and $e^-$. The collisions with the first three collision partners are charge exchange processes. Collisions with electrons may produce ENAs in radiative recombination. In search of the cross sections, we use the ALADDIN database maintained by the International Atomic Energy Agency[§].

### *4.1 Charge Exchange with Hydrogen Atoms ($H^+ + H^0 \rightarrow H^0 + H^+$)*

The charge exchange between protons and hydrogen atoms is the main process responsible for producing and losing hydrogen ENAs. This reaction is symmetric and resonant, i.e., no energy difference exists between the states before and after the collision. Consequently, the cross section for this process is large, even for low collision energy, unlike all other production processes.

The most frequently used formula to estimate this cross section was derived by Lindsay & Stebbings (2005) based on laboratory measurements. However, as discussed by Schultz et al. (2016), Bzowski & Heerikhuisen (2020), and Swaczyna et al. (2019b), this cross section may not be accurate in the whole the recommended range. We discuss these discrepancies and provide an analytic formula with recommended parameters to calculate this cross section in Appendix A based on theoretical calculations by Schultz et al. (2023) for collision energy up to 1 keV/u and recommended cross sections from the compilation by Barnett (1990) based on experimental results for higher collision energy. The cross section for this process monotonically decreases with collision energy from $4.9\times10^{-15}$ $cm^2$ at 1 eV/u to $5.1\times10^{-21}$ $cm^2$ at 500 keV/u. The cross section is $1.7\times10^{-15}$ $cm^2$ at the core solar wind energy of 1 keV/u (see Figure 1). The accuracy of the cross section is between 5% and 20%.

---

[§] https://www-amdis.iaea.org/ALADDIN/

### *4.2 Charge Exchange with Helium Atoms ($H^+ + He^0 \rightarrow H^0 + He^+$)*

The charge exchange of protons with helium atoms is another process producing ENAs. This process is often neglected because the density of helium atoms is lower in most regions of the heliosphere than the density of hydrogen atoms. Moreover, due to the difference in the ionization potentials of helium and hydrogen atoms, there is an energy defect reducing the cross section. Nevertheless, for higher collision energy, the helium cross sections exceed the one for the charge exchange with hydrogen atoms.

The ALADDIN database references several theoretical calculations, and a few derived cross sections based on laboratory measurements for this process. Barnett (1990, reaction A-32) is the most recent among the derived cross sections. The cross sections in Barnett (1990) are tabulated as coefficients of the Chebyshev polynomials up to the degree of eight. The polynomials use projectile energy transformed by applying the natural logarithm and normalized so that the applicable energy range is from 0 to 1. The sum of the polynomial values provides the natural logarithm of the recommended cross section. The applicability range for the cross section is from 99 eV/u to 11 MeV/u, covering almost the full range of the energies considered in our study. Moreover, for the lowest collision energies, the process is negligible. The largest cross section for this process is $\sim1.7\times10^{-16}$ $cm^2$ for the collision energy of ~24 keV/u. The uncertainty of this cross section is estimated at ~40% below 5 keV/u and ~20% above this threshold.

The collisions between protons and helium atoms may also produce hydrogen ENAs in the $H^+ + He^0 \rightarrow H^0 + He^{2+} + e^-$ process. The helium atom is fully ionized in this transfer ionization process, and the proton captures one electron. The cross section for this reaction increases for high energy collisions, but even then, the process is responsible for less than a few percent of the charge exchange process discussed above (Schmidt et al. 2005), and thus, we do not consider it further.

### *4.3 Charge Exchange with $He^+$ Ions ($H^+ + He^+ \rightarrow H^0 + He^{2+}$)*

Neutralization of protons in collisions with $He^+$ ions is also possible. The only cross section provided in the ALADDIN database is reaction A-54 from Barnett (1990). The recommended cross section covers collision energy from 1.8 to 240 keV/u. The largest cross section of $\sim2.5\times10^{-17}$ $cm^2$ is for the collision energy of ~41 keV/u. The reported accuracy is a factor of 2 below 10 keV/u and 25% above up to 220 keV/u. However, the uncertainty estimation for energy above 100 keV/u may be underestimated. Barnett (1990) relied on the experimental results from Rinn et al. (1985) and Angel et al. (1978a, 1978b), which do not agree for energies above 100 keV/u. Theoretical calculations of this cross section (e.g., Samanta et al. 2010; Faulkner et al. 2019) agree well with Rinn et al. (1985) but not with Angel et al. (1978a, 1978b).

Because we need to estimate the cross section for higher energies than the recommended range in Barnett (1990), we decide to use the results of theoretical calculations from Faulkner et al. (2019). We fit an analytic formula in the form used by Lindsay & Stebbings (1995) for the charge exchange between protons and hydrogen atoms to their results over the energy range from 50 keV/u to 1 MeV/u. The fit formula is:

$$\sigma(E) = (1.2375 + 0.12616 \log E)^2 [1 - \exp(-176.29/E)] \times 10^{-17}\ [\mathrm{cm}^2] \tag{10}$$

for energy $E$ in keV/u. This formula reproduces the theoretical cross section within ±4%. Moreover, Equation (10) agrees with the Barnett (1990) recommendation within ±10% for collision energies from 40 to 120 keV/u, but the discrepancy grows quickly for higher energies, and the Barnett (1990) cross section is 2.5 times larger at 240 keV/u. We recommend using the Barnett (1990) cross section for collision energies up to 100 keV/u and Equation (10) for higher energy. At 100 keV/u both formulae agree within 0.5%.

### *4.4 Radiative Recombination with Electrons ($H^+ + e^- \rightarrow H^0 + hv$)*

The recombination with electrons is the only available neutralization process for fully ionized plasma. The energy released in this process is emitted as electromagnetic radiation. Janev et al. (1987) provided the reaction rate coefficient formula, which we transformed into the cross section for comparison with other processes. The cross section is the largest for low energy and drops as the collision energy increases. Nevertheless, the process is negligible for typical relative energy in partially neutral space plasma.

### *4.5 Significance of ENA Production Processes*

The cross sections for the processes discussed in Sections 4.1-4.4 are shown in Figure 1. The highest cross section for low-energy collisions is for the charge exchange between protons and hydrogen atoms. However, around 50 keV/u, the cross section for the charge exchange with helium atoms becomes larger than that for the charge exchange with hydrogen above this energy. The cross section for the recombination with electrons is more than two orders of magnitude lower than the charge exchange cross section with hydrogen atoms.

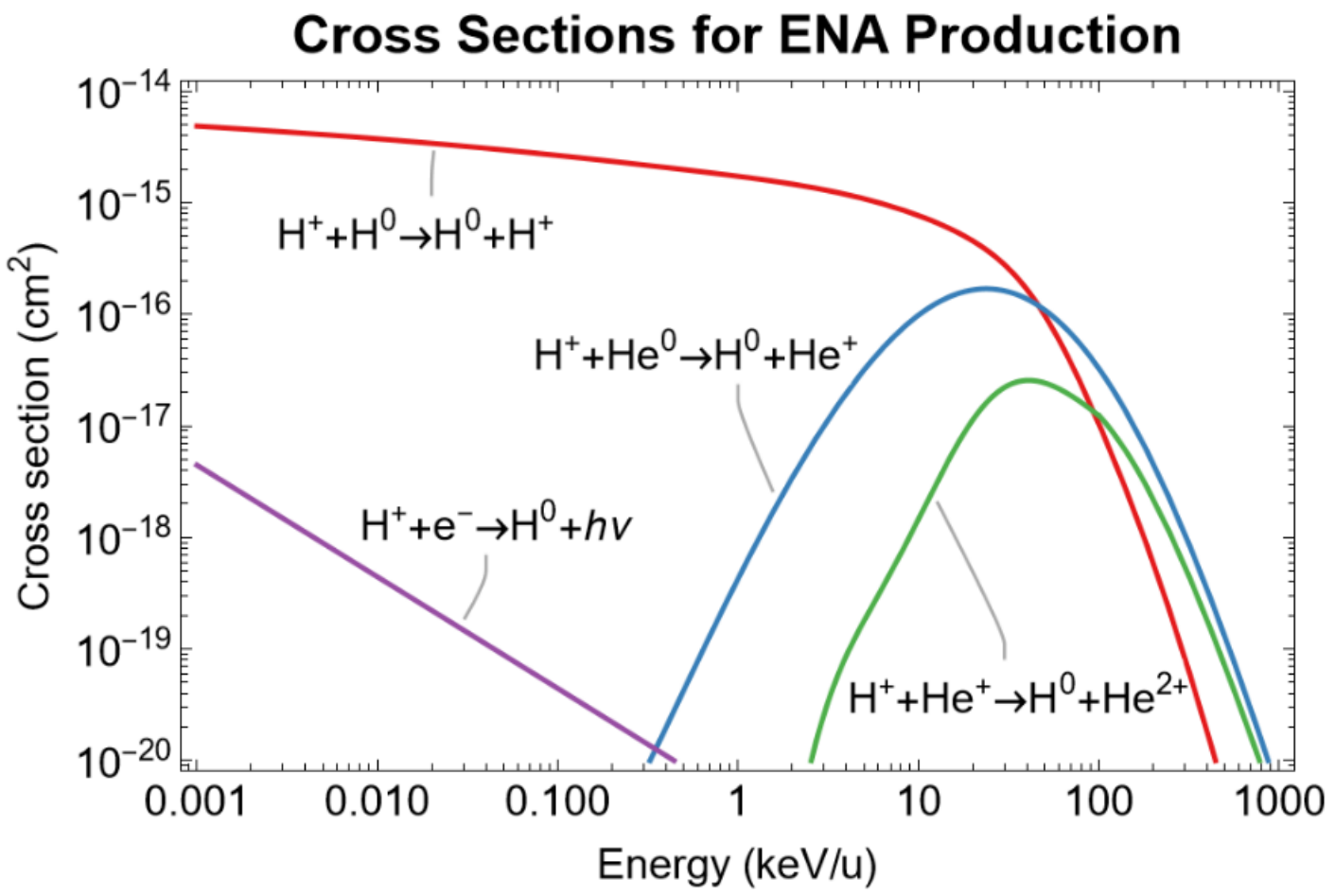

**Figure 1.** Comparison of the cross sections for the production of hydrogen ENAs. The cross sections are obtained from the formulae discussed in this paper, see Table 2 for references.

The cross section values are insufficient to judge the importance of each process because the collision rates further depend on the composition and relative flows of the collision partners. We use the six regions defined in Section 3 to estimate the collision rates and their relative importance. From the perspective of the ENA instruments, we are interested in the energy of ENAs in the solar-inertial frame. We neglect the momentum exchange in the ENA production processes, and therefore, the energy of the parent proton is the same as the ENA energy. Moreover, we assume that the parent proton moves radially inward to the Sun when it is neutralized.

For collisions with neutral components, we combine the parent proton speed with the bulk velocity of ISN populations and their thermal speeds "in squares" to estimate the collision speed. This calculation assumes that the ISN populations are moving perpendicularly to the radial direction, i.e., they correspond to the flanks of the heliosphere. Nevertheless, due to low ISN flow velocity, the impact of this approximation is negligible. For collisions with plasma components, i.e., electrons and $He^+$ ions, the ENA and bulk plasma velocities are assumed to be antiparallel in the supersonic solar wind and perpendicular in the heliosheath and VLISM. Note that these relations between the ENA energy are not valid for ENAs moving in other directions, e.g., for the primary ENAs in the secondary ENA mechanism.

The left column of Figure 2 shows the collision rates for each process calculated according to Equation (2) and the combined collision rate for all processes. The right column shows the relative contribution of each process to the combined collision rate of all processes. The figure compares the importance of each process over the energy range covering all three ENA instruments on IMAP.

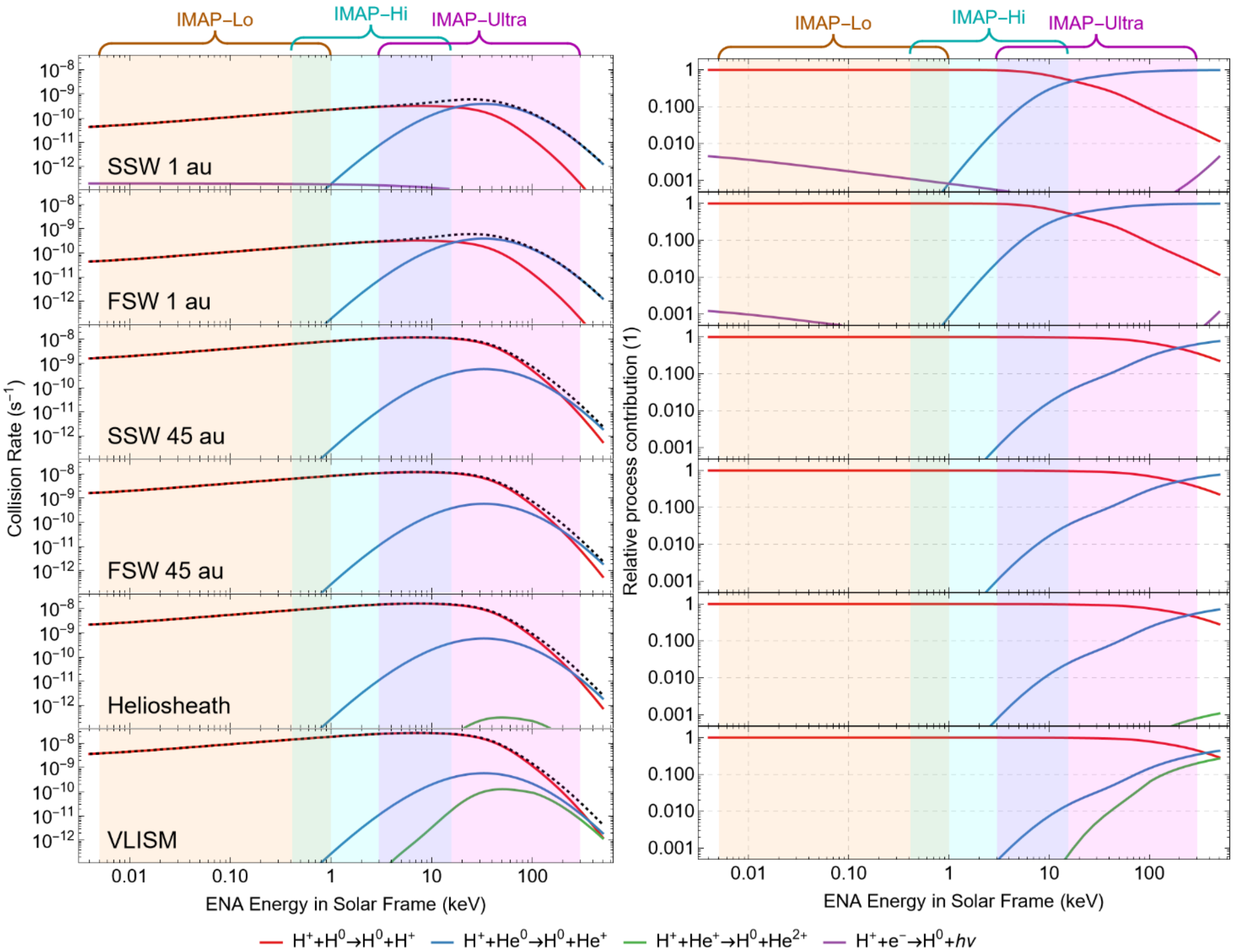

**Figure 2.** *Left column*: The collision rates for the ENA production processes in the heliospheric condition sets defined in Table 1 (color lines) and the combined production rate (black dotted line). *Right column*: the relative contributions of individual processes to the combined collision rate. The shaded areas correspond to the energy ranges of the IMAP ENA instruments: IMAP-Lo (orange), IMAP-Hi (cyan), and IMAP-Ultra (magenta). Note the energy ranges overlap between the instruments.

We use three categories to classify the processes based on the aimed accuracy of ENA flux calculations. The reactions contributing at least 10% or 1% to the production rate are classified in categories A or B, respectively. The other reactions are included in category C. We report the highest category for each IMAP instrument over the respective energy range. Table 2 provides all categories if the assigned category depends on the region considered.

The charge exchange with hydrogen atoms is a category A process for all three instruments. In the IMAP-Hi energy range, the charge exchange with helium is a category A process for conditions at 1 au due to lower ISN hydrogen density and a category B process for other sets. For the IMAP-Ultra range, the charge exchange with helium is always a category A process. Moreover, in the VLISM, the charge exchange with $He^+$ is a category A process, while it is a category C process elsewhere. At the high-energy boundary of the applicability of this cross section, the process may exceed the charge exchange with hydrogen atoms. The recombination with electrons is negligible for all cases (category C).

Based on the above assessment, the studies considering the ENA production within the IMAP-Lo energy range may focus on the charge exchange between protons and hydrogen atoms. The charge exchange with helium atoms may contribute to the IMAP-Hi energy range, especially for the highest energies. For IMAP-Ultra, the charge exchange with both neutral species should be included and the charge exchange with $He^+$ ions in $He^+$-rich regions (e.g., in the VLISM).

**Table 2. Hydrogen ENA Production Reactions**

| | | | Significance category | | |
|---|---|---|---|---|---|
| **Reaction** | **Section** | **Cross Section Source** | **IMAP-Lo** | **IMAP-Hi** | **IMAP-Ultra** |
| $\mathbf{H^+ + H^0 \rightarrow H^0 + H^+}$ | 4.1 | Appendix A, Equation (A1) | A | A | A |
| $\mathbf{H^+ + He^0 \rightarrow H^0 + He^+}$ | 4.2 | Barnett (1990, A-32) | C | A/B | A |
| $\mathbf{H^+ + He^+ \rightarrow H^0 + He^{2+}}$ | 4.3 | Barnett (1990, A-54) & Equation (10) | C | C | A/C |
| $\mathbf{H^+ + e^- \rightarrow H^0 + \mathit{h\nu}}$ | 4.4 | Janev et al. (1987, Section 2.1.8) | C | C | C |

## 5. Losses of Hydrogen ENAs

There are two groups of ENA loss processes. The first group consists of charge exchange reactions, in which an electron transfers between the colliding particles. In this group, we consider collisions with atoms and ions. The other group consists of ionization processes in which an electron is stripped from an ENA and freely emitted. The ionization is possible in collisions with ions, atoms, electrons, and photons. The charge exchange processes are important for low collision energies, and the ionization processes become important for higher energies.

### *5.1 Charge Exchange with Protons ($H^0 + H^+ \rightarrow H^+ + H^0$)*

The charge exchange with protons is discussed in Section 4.1. While the number of hydrogen atoms as a result of the process does not change, due to the difference in the parent ion properties, the newly created ENA may move in a completely different direction. This process is the main loss process at low and mid ENA energy.

### 5.2 *Charge Exchange with Hydrogen Atoms ($H^0 + H^0 \rightarrow H^+ + H^-$)*

Collisions between two hydrogen atoms may result in electron exchange, producing protons and hydrogen anions ($H^-$). The ALADDIN database recommends the cross section from Barnett (1990, reaction A-2) for collision energy from 2 to 73 keV/u. The maximum cross section for this process is ~$1.6\times10^{-17}$ $cm^2$ for the collision energy of ~17 keV/u. The cross section drops by a factor of ~10 at the ends of the applicability range compared to the maximum. The accuracy of this cross section is estimated at 50% below 8 keV/u and 20% above this threshold. We note that the cross section calculated from the Chebyshev polynomial coefficients given by Barnett (1990) has a distinctive "hump" around 2.5 keV/u, which is not visible in the recommended data.

The process involves two hydrogen atoms. If the considered ENA turns into a proton, the loss is like the other considered loss processes. However, the situation is different if the considered ENA turns into a hydrogen anion. The electron affinity in hydrogen anions is only ~0.754 eV. Hence, hydrogen anions quickly lose electrons by photodetachment. The photodetachment rate at 1 au is ~14.3 $s^{-1}$ and decreases as $r^{-2}$ with distance $r$ from the Sun (Desai et al. 2021). Therefore, the mean lifetime at 1 au is ~0.07 s and increases to ~700 s at 100 au from the Sun. We compare the lifetime to the gyroperiod of the anions in the solar wind. Assuming the interplanetary magnetic field of 40 μG at 1 au, which, according to the Parker spiral, decreases as $\sim r^{-1}$, the gyroperiod decreases from ~13 s at 1 au to ~1300 s at 100 au. Therefore, near 1 au, the hydrogen anions turn back into ENAs after only ~0.005 of the gyroperiod. While the ENAs formed from these anions have a slightly different direction, they continue the journey from where the original ENA was ionized. Therefore, we do not consider this process as a loss source. However, the process "blurs" the observed ENA structures. Even though the process could impact the ENA fluxes in a unique way, the impact on the global ENA maps is negligible because this process is insignificant (see Section 5.13) compared to other ionization sources.

### 5.3 *Charge Exchange with α-particles ($H^0 + He^{2+} \rightarrow H^+ + He^+$)*

The charge exchange with α-particles is another ENA loss source. Barnett (1990, reaction A-88) and Janev & Smith (1993, Section 3.3.1) provided formulae based on experimental measurements and theoretical calculations. Both formulae are in good agreement for collision energy above 1 keV/u, the Janev & Smith (1993) formula reproduces better the Havener et al. (2005) theoretical calculations for collision energy down to 380 eV/u. Therefore, we decide to use their formula in our study.

The formula can be applied to energy from 100 eV/u to 1 MeV/u. The maximum cross section of $1.2\times10^{-15}$ $cm^2$ is for the energy of ~10 keV/u. The cross section drops by more than a factor of 1000 at the edges of the energy range. The accuracy is estimated to be ~20-60% from 0.1 keV/u to 0.6 keV/u and 20% above 600 eV/u. Nevertheless, we use the analytic formula also for energies below the recommended range.

### 5.4 *Charge Exchange with $He^+$ Ions ($H^0 + He^+ \rightarrow H^+ + He^0$)*

The next considered process is the charge exchange with $He^+$ ions. The only source of the cross section cited in the ALADDIN database is Barnett (1990, reaction A-62). The maximum cross section for this charge exchange is $2.6\times10^{-16}$ $cm^2$ at ~16 keV/u. The formula can be applied for energy from 0.25 keV/u to 700 keV/u. The estimated accuracy of this cross section is 30% above 0.6 keV/u and unknown below this threshold. The cross section is $1.5\times10^{-17}$ $cm^2$ and $2.5\times10^{-21}$ $cm^2$ at the low and high ends of the energy range.

Recent theoretical calculations by Gao et al. (2024) cite the same experiments as used by Barnett (1990) and show a reasonable agreement between the theoretical calculations and the experimental data for higher energies. Nevertheless, the low-energy end of the Barnett (1990) cross section used the theoretical calculations and is within 50% of the Gao et al. (2024) calculations. However, the process is unimportant for ENA losses at these energies, and the Barnett (1990) cross section can be used.

### *5.5 Charge Exchange with Helium Atoms ($H^0 + He^0 \rightarrow H^- + He^+$)*

Collisions of ENAs with helium atoms may also result in a charge exchange producing $H^-$ and $He^+$ ions. The cross section for this process is given as reaction A-10 in Barnett (1990), but for the reason discussed in Section 5.2, we do not consider the production of hydrogen anions as a loss term due to the short lifetime of such ions. Additionally, the cross section for this process is below $10^{-17}$ cm$^2$, which is at least by a factor of 10 less than the ionization cross section discussed in Section 5.12.

### *5.6 Photoionization ($H^0 + hv \rightarrow H^+ + e^-$)*

Photoionization is an important ENA loss process in the heliosphere. The energy required to ionize a hydrogen ENA is 13.6 eV, corresponding to the wavelength of 91.18 nm. Only photons with wavelengths below this threshold can ionize hydrogen ENAs. The analytic formula for the photoionization cross section is given by Verner et al. (1996). The maximum cross section value is ~$6.3\times10^{-18}$ cm$^2$ at the threshold energy. However, the solar irradiance in the extreme UV range is complex, with multiple emission lines dominating the continuous spectrum (de Wit et al. 2005). Due to the decrease in the cross section, the main part of the photoionization is from the wavelengths just below the threshold, although the 30.4 nm $He^+$ line contributes to this process.

The photoionization rate at 1 au is estimated based on the observations of solar irradiance by TIMED/SEE (Woods et al. 2005). Integrating the cross section with the solar spectrum provides the photoionization rate ranging from $0.8\times10^{-7}$ s$^{-1}$ to $1.4\times10^{-7}$ s$^{-1}$ at 1 au (Bochsler et al. 2014). The solar radiation flux decreases proportionally to $r^{-2}$. The photoionization rate decreases proportionally to this flux.

For the broad range of energy of ENAs considered in this study, we considered the Doppler shift of the solar spectrum in the rest frame of the atom. For the highest considered energy (500 keV), the atom's speed is about 0.033$c$, where $c$ is the speed of light. The relativistic Doppler effect shifts the wavelengths in the atom's frame by a factor of $\sqrt{\frac{1-v/c}{1+v/c}} \approx 0.968$ if the atom moves radially toward the Sun. Nevertheless, integrating the Doppler shifted spectrum with the cross section increases the photoionization rate by only about 0.6%. Therefore, we do not consider this change further in our analysis, and we treat the photoionization rates as independent of the ENA energy. We use the mean photoionization of ~$1.1\times10^{-7}$ s$^{-1}$ at 1 au in the analysis.

### *5.7 Electron impact ionization ($H^0 + e^- \rightarrow H^+ + 2e^-$)*

Electron impact ionization is another source of ENA losses in the heliosphere. The cross section recommended in the ALADDIN database is from Janev & Smith (1993, Section 1.2.1). The minimum electron energy in the ENA rest frame needed for this ionization equals the electron's binding energy in the hydrogen atom, i.e., 13.6 eV. The maximum cross section is $6.2\times10^{-17}$ cm$^2$ for the electron energy of ~57.2 eV. The cross section rapidly decreases below this energy. This cross section is lower by ~5% near the maximum than the formula given by Lotz (1967). However, the Janev & Smith (1993) results are consistent with high-accuracy experimental results (Shah et al. 1987), unlike the Lotz (1967) formula.

Because the thermal velocities of electrons are comparable with the speeds of considered ENAs, the approximation given by Equations (6-7) may be inaccurate. Thus, we numerically integrate the formula in Equations (4-5) for wide ranges of electron temperatures and relative speeds of atoms in the plasma frame. The results for the temperatures from $10^3$ to $10^8$ K and velocities corresponding to ENA energy from 5 eV to 500 keV are shown in Figure 3. We also calculate the reaction rate coefficient using the approximation from Equations (6-7).

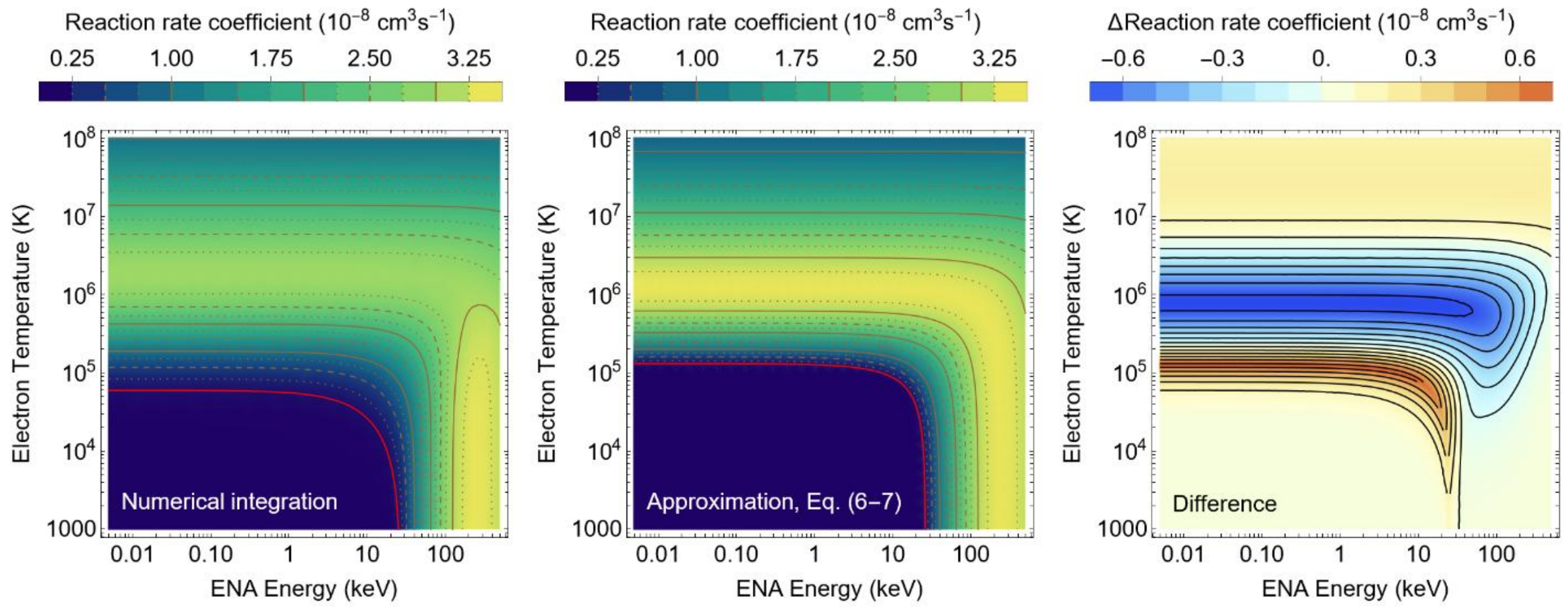

**Figure 3.** Reaction rate coefficient for the electron impact ionization of hydrogen ENAs. The left and middle panels show the results of the numerical integration with the Maxwellian distribution, and the approximation given in Equations (6-7), respectively. The right panel presents the difference between the calculated rate coefficients using these two approaches. The red lines in the left and middle panels mark the reaction rate coefficient of $10^{-9}$ cm$^3$ s$^{-1}$.

The comparison indicates that the approximation utilizing the mean relative speed is inaccurate for low-energy ENAs. The approximation overestimates the reaction rate for temperatures around $10^6$ K because the actual cross section calculated for different portions of the distribution function is typically lower than that calculated for the mean relative speed. The overestimation may reach ~20%. On the other hand, the reaction rate is underestimated for temperatures ~$10^5$ K, for which the mean relative speed corresponds to the threshold ionization energy. In this case, the ionization source is mostly from a portion of the Maxwell distribution, for which the relative speed is higher. In the region where the difference is the highest in the right panel of Figure 3, the reaction rate coefficient is underestimated by a factor of ~5 (compare the position of the red line in the middle panel with the dashed line near the same place in the left panel), or even more for lower temperatures. The comparison shows that the numerical calculation of the rate coefficient is essential for ENA energy up to ~30 keV/u and electron temperatures of ~$10^5$ K.

### *5.8 Ionization by Protons ($H^0 + H^+ \rightarrow H^+ + H^+ + e^-$)*

Collisions of ENAs with ambient particles in the interplanetary and interstellar space may result in stripping electrons. This and the four next sections consider such reactions. The ALADDIN database identifies two sources of analytic cross sections for ionization by protons: Barnett (1990, reaction D-6) and Janev & Smith (1993, Section 2.2.1). Both formulae agree within a maximum difference of 7% over the common applicability range. We select Janev & Smith (1993) due to the broader energy range (from 0.2 keV/u to 4 MeV/u), which is needed especially for low-energy collision. The estimated accuracy of the cross section is between 10% and 30% below 10 keV/u and better than 10% above 10 keV/u. The maximum cross section is ~$1.4\times10^{-16}$ $cm^2$ for the collision energy of ~53 keV/u. The cross section drops to ~$10^{-19}$ $cm^2$ near the collision energy of 1 keV/u and ~$10^{-17}$ $cm^2$ near 2 MeV/u. While the Janev & Smith (1993) formula is consistent with the experimental data, some newer theoretical approaches predict a slightly higher cross section, especially near the peak of the cross section (see discussion in Cariatore & Schultz 2021).

### *5.9 Ionization by Hydrogen Atoms ($H^0 + H^0 \rightarrow H^+ + H^0 + e^-$)*

Hydrogen atoms are the most abundant species in the outer heliosphere and VLISM. Therefore, the ionization by collision with these atoms may be an important source of losses. The only cross section source listed by ALADDIN is Barnett (1990, reaction E-2). The Chebyshev polynomial provided in this source is applicable for the energy range from 1.2 keV/u to 3.5 MeV/u, but the cross section is still significant ($3\times10^{-18}$ $cm^2$) at the lower boundary. Moreover, several sources of experimental data and theoretical calculations were not included (Gealy & Van Zyl 1987; Soon 1992; Ovchinnikov et al. 2017). Cariatore & Schultz (2021) provided a recommended cross section including this additional source in a numerical form for the energy range from 30 eV/u to 10 MeV/u. Appendix B provides an analytic formula that fits their recommendation that we use in our study. The largest cross section is ~$1\times10^{-16}$ $cm^2$ for the collision energy of ~23 keV/u. The cross section drops to ~$8\times10^{-22}$ $cm^2$ and ~$1.4\times10^{-18}$ $cm^2$ at the lower and upper end of the recommended range, respectively. Considering the accuracy of Barnett (1990) and uncertainties of the Gealy & Van Zyl (1987) results, the cross section is known with an accuracy of ~20-30%.

Cariatore & Schultz (2021) also considered simultaneous ionization and stripping of colliding partners in a collision of hydrogen atoms, i.e., that both colliding atoms eject electrons. While such a process would also be an ENA loss process, the cross section for this process is at least 12 times smaller than the process discussed here, i.e., it is much smaller than the uncertainty.

### *5.10 Ionization by α-particles ($H^0 + He^{2+} \rightarrow H^+ + He^{2+} + e^-$)*

As for collisions with protons, the ALADDIN database provides two analytic formulae for ionization by α-particles: Barnett (1990, reaction D-96) and Janev & Smith (1993, Section 3.2.1). We use the Janev & Smith (1993) formula because this analysis included experimental results for lower energies from Shah et al. (1988) and provides cross section for a broader energy range. The maximum cross section is $4.5\times10^{-16}$ $cm^2$ for the collision energy of ~77 keV/u. The applicability range of the formula is from ~0.5 keV/u to 5 MeV/u. The accuracy is low (~100%) near the lower end of this range but increases to 10-20% at 10 keV/u and is better than 10% above 30 keV/u. The cross section is ~$10^{-20}$ $cm^2$ at 1 keV/u and $8\times10^{-17}$ $cm^2$ at 1 MeV/u.

### *5.11 Ionization by $He^+$ Ions ($H^0 + He^+ \rightarrow H^+ + He^+ + e^-$)*

The ALADDIN database cites only the Barnett (1990, reaction D-72) cross section for the electron stripping by $He^+$ ions. The maximum cross section is ~$2.1\times10^{-16}$ $cm^2$ for the collision energy of ~62 keV/u. The provided formula should be used for collision energy from 6.3 keV/u to 8 MeV/u, for which the cross section values are $1.3\times10^{-17}$ $cm^2$ and $4.5\times10^{-18}$ $cm^2$, respectively. The formula was derived based on theoretical calculations because there had not been any experimental results for this reaction. The only experimental results we found are from Hsu et al. (1996), which provide results for 5 collision energies from 7 keV/u to 28.5 keV/u. The experimental cross sections are 40%-75% higher than those calculated from the Barnett (1990) formula. Nevertheless, considering the narrow energy range of the experimental data, we recommend using the Barnett (1990) formula to calculate hydrogen ENA losses.

### *5.12 Ionization by Helium Atoms ($H^0 + He^0 \rightarrow H^+ + He^0 + e^-$)*

The last hydrogen ENA loss source considered here is the stripping by helium atoms. The ALADDIN database refers to Barnett (1990, reaction E-10) for this process. The cross section is derived based on some theoretical and experimental data. The maximum value is ~$1.4\times10^{-16}$ cm$^2$ at 5.8 keV/u. The applicable energy range is from 50 eV/u to 2 MeV/u, for which the cross section value is ~$2\times10^{-19}$ cm$^2$ and $7\times10^{-19}$ cm$^2$. The estimated accuracy of the formula is 15%.

### *5.13 Significance of ENA Loss Processes*

We consider 11 ENA loss processes (Sections 5.1-5.12 except 5.5). Figure 4 compares the cross sections for all of them except the photoionization. For low collision energies, the only important ENA loss process is the charge exchange with protons. At mid-energies, charge exchange processes with α-particles exceed the cross section for charge exchange with protons. The charge exchange cross sections with He$^+$ ions and hydrogen atoms are approximately 10 and 100 times smaller, respectively. Finally, for the highest energies considered in this paper, a number of ionization reactions in collisions with various species dominate . For the collisions with charged ions, the charge exchange cross sections drop rapidly above a few tens of keV/u. Nevertheless, the ionization process cross sections increase when the charge exchange cross sections drop. Therefore, the combined losses due to these ions do not change as rapidly as one may expect from the charge exchange cross sections alone.

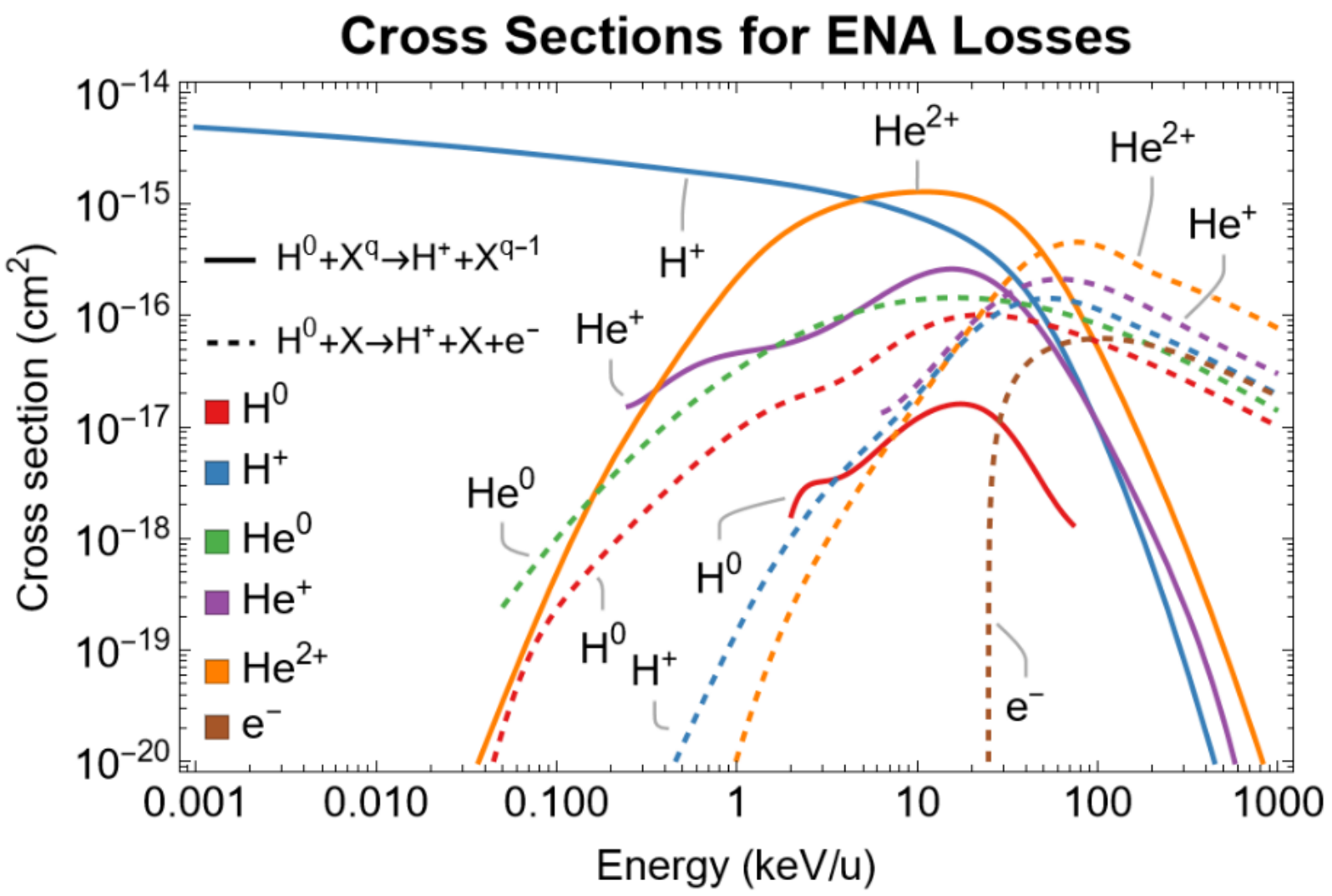

**Figure 4.** Cross sections for ENA loss processes discussed in Section 5 (except photoionization) as a function of the collision energy. The cross sections for the charge exchange and ionization losses in collisions with the same species are marked with solid and dashed lines of the same color.

We need to consider the densities and temperatures in different regions of the heliosphere to evaluate the importance of each process fully. For this purpose, we calculate the collision rates in each region, which express the probability of ENA loss per unit time. The rate is calculated assuming that the ENA energy is in the solar-inertial frame. Note that this energy does not directly correspond to the collision energy shown in Figure 4. The collision rates are presented in the left panel of Figure 5. At 1 au, the main reaction is the charge exchange with protons (Section 5.1). However, while the cross section for this process decreases above ~20 keV, the collisional ionization with protons (Section 5.8) balances this drop. Finally, the combined rate of the processes is approximately constant through the energy range.

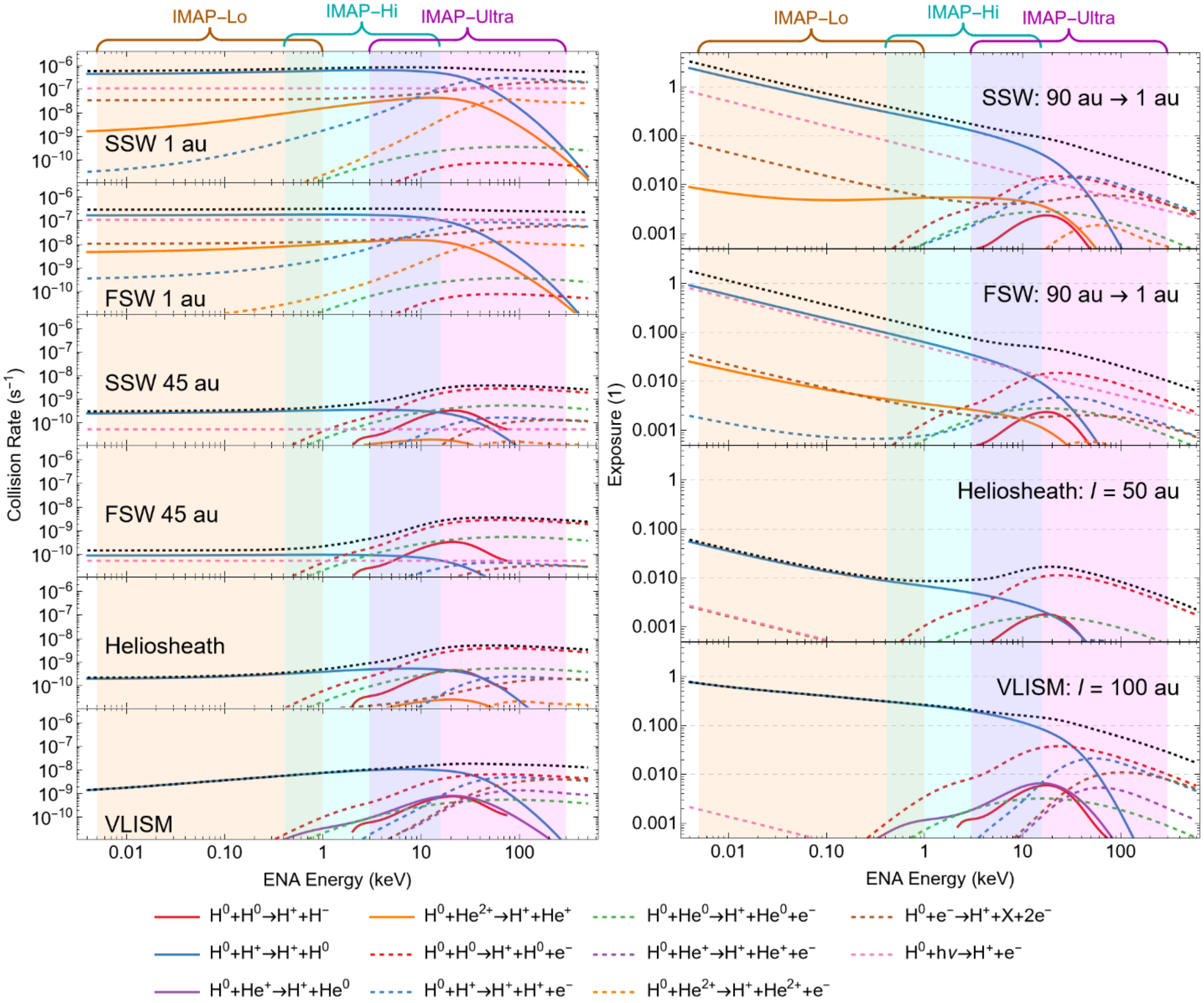

**Figure 5.** *Left column*: The collision rates for the ENA loss processes in the heliospheric condition sets defined in Table 1 (color solid and dashed lines) and the combined collision rate (black dotted line). *Right column*: The ENA loss exposure factors for ENAs traversing the SSW, FSW, heliosheath, and VLISM regions (panels top to bottom). The line colors and styles are the same, as shown in Figure 4. The colors of shaded areas correspond to IMAP ENA detector energy ranges, as in Figure 2.

The situation is slightly different at 45 au from the Sun, where the interstellar hydrogen density is higher, and thus, the collisional ionization by hydrogen atoms (Section 5.9) becomes the main loss reaction above ~5-10 keV. The situation in the heliosheath is similar. In the VLISM, the charge exchange with protons is the main source of losses of <~20 keV ENAs, while ionization by hydrogen atoms, protons, and electrons dominate the losses of more energetic atoms. The processes operating in the VLISM limit the distance range that can be probed using hydrogen ENAs.

The ENA losses from the source to a detector at 1 au caused by solar wind protons, electrons, and photoionization dominate for low energy ENAs, and thus, the collision rate drops with distance, following the solar wind density. However, for high-energy ENAs, the ionization by hydrogen atoms, which are present everywhere in the heliosphere, is spread more uniformly along the path of ENAs.

We do not judge the importance of each process based on the relative contribution to all ionization losses but rather on the relative loss caused by each process. Therefore, we estimate the exposure factor calculated from Equation (8) for ENA (a) traversing the supersonic solar wind radially from 90 au to 1 au, (b) traveling 50 au in the heliosheath, and (c) traveling 100 au in the VLISM. For the heliosheath and VLISM, we assume that the conditions are homogenous along the path with the densities and temperatures as described in Section 3.

For the supersonic solar wind, we estimate the parameter change following the rules discussed in Section 3 to derive the conditions at 45 au with the following exceptions. The thermal speeds of solar wind protons and helium ions are 20 km $s^{-1}$ and 10 km $s^{-1}$, respectively. While the plasma temperature changes with distance from the Sun, the thermal speeds of these particles are typically negligible compared to the ENA speeds, and a detailed model is not necessary. The PUI thermal speeds are assumed constant as in the SSW and FSW sets. However, the electron thermal speeds are comparable with the ENA speeds. Therefore, we assume that the temperatures of the two considered populations change with distance as $T_c = 10^5 r^{-0.85} + 10^4$ K and $T_h = 10^6 r^{-0.38} + 10^4$ K, for the $r$ in au. The power-law indices are adopted as in Bzowski et al. (2013). The added $10^4$ K ensures that the electron temperature is comparable to the proton temperature in the distant solar wind, in other words, the temperature tends asymptotically to $10^4$ K.

The estimated ENA loss exposure factors are presented in the right column of Figure 5. The combined exposure is the highest for the lowest ENA energy, whereas, in the supersonic solar wind (SSW and FSW), it may exceed 1. The main factors considered in the past (Bzowski et al. 2013), i.e., the charge exchange with protons and α-particles, electron impact ionization, and photoionization, are the only relevant loss processes for ENA energy up to a few keV. However, several ionization processes contribute to ENA losses for higher energy.

We assign the importance categories for the loss reactions following the criteria listed in Section 4.5 based on the exposure in the supersonic solar wind (see Table 3). The category A and B processes are those for which the exposure exceeds 0.1 and 0.01, respectively, for any energy within the instrument's energy range. The charge exchange with protons is a category A process, except for the IMAP-Ultra energy range in the FSW, for which it is a category B process. The photoionization is a category A process in the IMAP-Lo range, but only a category B process for IMAP-Hi and IMAP-Ultra. Other category B processes are ionization in collisions with hydrogen atoms (IMAP-Hi and IMAP-Ultra) and electron impact ionization (IMAP-Lo). Finally, the charge exchange with α-particles is a category B process in the FSW (IMAP-Lo), and the ionization by protons is a category B process in the SSW (IMAP-Ultra). All other processes contribute less than 0.01 to the exposure and are considered category C.

**Table 3. Loss Reactions of Hydrogen ENAs**

| Reaction | Section | Cross Section Source | Significance category | | |
|---|---|---|---|---|---|
| | | | IMAP-Lo | IMAP-Hi | IMAP-Ultra |
| $\mathbf{H^0 + H^+ \rightarrow H^+ + H^0}$ | 5.1 | Appendix A, Equation (A1) | A | A | A/B |
| $\mathbf{H^0 + \mathit{h\nu} \rightarrow H^+ + e^-}$ | 5.6 | Verner et al. (1996) | A | B | B |
| $\mathbf{H^0 + H^0 \rightarrow H^+ + H^0 + e^-}$ | 5.9 | Cariatore and Schultz (2020, Table 3), see also Appendix B | C | B | B |
| $\mathbf{H^0 + e^- \rightarrow H^+ + e^- + e^-}$ | 5.7 | Janev & Smith (1993, Section 1.2.1) | B | C | C |
| $\mathbf{H^0 + He^{2+} \rightarrow H^+ + He^+}$ | 5.3 | Janev & Smith (1993, Section 3.3.1) | B/C | C | C |
| $\mathbf{H^0 + H^+ \rightarrow H^+ + H^+ + e^-}$ | 5.8 | Janev & Smith (1993, Section 2.2.1) | C | C | B/C |
| $\mathbf{H^0 + H^0 \rightarrow H^+ + H^-}$ | 5.2 | Barnett (1990, reaction A-2) | C | C | C |
| $\mathbf{H^0 + He^+ \rightarrow H^+ + He^0}$ | 5.4 | Barnett (1990, reaction A-62) | C | C | C |
| $\mathbf{H^0 + He^{2+} \rightarrow H^+ + He^{2+} + e^-}$ | 5.10 | Janev & Smith (1993, Section 3.2.1) | C | C | C |
| $\mathbf{H^0 + He^+ \rightarrow H^+ + He^+ + e^-}$ | 5.11 | Barnett (1990, reaction D-72) | C | C | C |
| $\mathbf{H^0 + He^0 \rightarrow H^+ + He^0 + e^-}$ | 5.12 | Barnett (1990, reaction E-10) | C | C | C |

The role of ionization processes in the inner heliosheath is limited. The exposure factor is from one to a few percent over the distance of 50 au. Because the production of ENAs is limited by the depopulation of energetic protons in distant regions of the heliosphere (Schwadron et al. 2014), most of the ENAs are produced within ~100-200 au from the Sun. Still, the losses due to charge exchange with protons and ionization by hydrogen atoms may slightly reduce the fluxes of the most distant source regions in the heliosheath.

The situation is different in the VLISM, where the charge exchange with protons reduces the fluxes by more than 10% up to the energy of ~30 keV over a distance of 100 au. Therefore, ENA sources in the VLISM can be observed only within a few hundred au from the Sun. Nevertheless, observations of helium ENAs may allow for observations of regions with considerable amounts of energetic ions up to ~10,000 au from the Sun (Swaczyna et al. 2017).

## 6. Summary

The IMAP ENA instruments will cover a broad energy range of hydrogen ENAs from 5 eV to 300 keV. We identified the binary collisions occurring in outer space involving charged particles (electrons, photons, protons, $He^+$ ions, α-particles) and neutral atoms (hydrogen and helium) that may produce or destroy hydrogen ENAs. We analyzed the importance of these processes in various physical conditions representative of different parts of the heliosphere.

We identified four possible ENA production processes: charge exchange of energetic protons with hydrogen atoms, helium atoms, $He^+$ ions, and radiative recombination with electrons. The charge exchange of energetic protons with hydrogen atoms is the main source of hydrogen ENAs for low collision energies. However, at ~10 keV/u, the charge exchange with helium atoms produces a few percent of hydrogen ENAs. Therefore, modeling hydrogen ENA fluxes at the highest energy steps of IMAP-Hi and throughout the entire energy range of IMAP-Ultra should account for this process. Above ~200 keV/u, this process may dominate over the charge exchange with hydrogen.

The charge exchange with $He^+$ ions may further contribute to ENA production above ~100 keV/u in the regions where the density of $He^+$ ions is significant, e.g., in the VLISM. Therefore, modeling of the hydrogen ENA emission from beyond the heliopause in the highest ENA energies must consider all these three charge exchange processes. However, we confirmed that the radiative recombination is negligible in all considered physical conditions.

ENAs are ionized on their path from the source region by many processes. We evaluated the importance of four charge exchange and seven ionization processes in this respect. The main loss process for the IMAP-Lo energy range is the charge exchange with protons. Additionally, the photoionization ionizes up to a similar amount of ENAs, while the electron impact ionization contributes a few percent of the losses. While mostly the same processes ionize hydrogen ENAs in the IMAP-Hi energy range, the ionization in collisions with neutral hydrogen decreases the ENA flux by only a few percent above ~10 keV. Finally, the main loss processes in the IMAP-Ultra energy range above 30 keV are the ionization collisions with hydrogen atoms, protons, and electrons.

The ionization losses in the heliosheath attenuate the ENA flux by at most a few percent over a distance of 50 au (see Figure 5). Therefore, the observations of ENAs from the heliosheath are mostly limited by the cooling length of energetic protons in the heliosheath (Schwadron et al. 2011, 2014; Kornbleuth et al. 2023) rather than by the ionization losses. However, any extraheliospheric sources of ENAs are attenuated over just a few hundred au in the interstellar medium.

## Acknowledgments

The authors thank the anonymous reviewer for providing their preliminary results of theoretical calculations of the cross section discussed in Section 4.3. P.S. is supported by a project co-financed by the Polish National Agency for Academic Exchange within the Polish Returns Programme (BPN/PPO/2022/1/00017) with a research component funded by the National Science Centre, Poland (2023/02/1/ST9/00004). M.B. and M.A.K. are supported by the National Science Centre, Poland (2023/51/B/ST9/01921). 

## Appendix A. Cross section for $H^0 + H^+ \rightarrow H^+ + H^0$

Lindsy & Stebbings (2005) derived a broadly used analytic formula to calculate the charge exchange cross section for collision energy from 5 eV to 250 keV. Their derivation relied on multiple experimental results collected from 1960 to 1987. The list of data sources used in their paper is a subset of the data source used by Barnett (1990, reaction A-22). The sources in the Barnett (1990) analysis excluded by Lindsay & Stebbings (2005) are not only experiments for collision energies above the upper limit of the Lindsay & Stebbings (2005) recommended range (Hvelplund & Andersen 1982; Schwab et al. 1987) but also experiments within this range (Fite et al. 1958; Gilbody & Ryding 1966; Bayfield 1969). Lindsy & Stebbings (2005) did not discuss the omission of these data sources in their paper. Therefore, even though their compilation is more recent, it is not necessarily better than the Barnett (1990) recommendation.

Recently, Bzowski & Heerikhuisen (2020) noticed that the Linsday & Stebbings (2005) cross section is up to ~40% larger than the cross section calculated according to the Chebyshev polynomial given by Barnett (1990) for the low collision energies. The discrepancy is likely because Linsday & Stebbings (2005) excluded laboratory data based on collisions with deuterium from Newman et al. (1982), even though the deuterium collision results could be scaled to the protium collisions.

The measurements at low collision speeds are complicated for two reasons. First, the collision energy is hard to control. Furthermore, the momentum exchange is no longer negligible. Thus, identifying the collision products is problematic (Schultz et al. 2008). On the other hand, the theoretical calculations provide accurate results in low-energy regimes. For these reasons, the ALADDIN database assigned a high accuracy grade to the theoretical cross sections at low collision energy up to ~200 eV. Schultz et al. (2016) performed comprehensive calculations of the cross section for CM energy from $10^{-4}$ to $10^4$ eV and found that the Lindsay & Stebbings (2005) formula overestimates their results by up to ~20% for projectile energy of a few eV. In other words, the Schultz et al. (2016) cross section is approximately halfway between the Lindsay & Stebbings (2005) and Barnett (1990) cross sections. They found a good match for the projectile energy around 1 keV/u, but a significant discrepancy is seen for energy ~10 keV/u due to limitations of the accuracy of their calculation methodology in this range.

Swaczyna et al. (2019b) proposed a new set of parameters for the analytic form suggested by Lindsay & Stebbings (2005) that provide a good fit to the predictions of the theoretical calculations of Schultz et al. (2016) below the projectile energy of 0.2 keV and reproduce the Lindsay & Stebbings (2005) formula within ±10% between 0.2 keV and 250 keV. This updated formula reproduces the theoretical predictions within ±10% for the projectile energy from 0.02 eV to 6 keV. The discrepancy between the formulae increases above 250 keV, but this is beyond the recommended energy range of Lindsay & Stebbings (2005).

In addition to the discrepancy at the low collision energies, the cross sections from Lindsay & Stebbings (2005) and Barnett (1990) differ for collision energy ~20 keV/u, which propagated to the formula found by Swaczyna et al. (2019b). Therefore, we find a new approximation formula that relies on theoretical calculations from Schultz et al. (2023) for collision energy below 1 keV/u and uses the recommended cross sections from Barnett (1990) for higher energies. Unfortunately, updating parameters in the Lindsay & Stebbings (2005) formula does not produce satisfactory results because the formula has only three free parameters, which cannot correctly capture the complexity of the cross section.

Janev & Smith (1993, Section 2.3.1) proposed the following analytic form for this process:

$$\sigma(E) = 10^{-16} \frac{A_1 \ln(A_2/E + A_3)}{1 + A_4 E + A_5 E^{A_6} + A_7 E^{A_8}}, \tag{A1}$$

where $\sigma(E)$ is the cross section in cm$^2$ for $E$ in keV/u, and $A_1$-$A_8$ are the fit parameters. They found fit parameters using the Barnett (1990) recommended cross sections for collision energy below 400 keV/u. In our fit, we use Schultz et al. (2023) theoretical calculations from 0.1 eV/u to 1 keV/u and the Barnett (1990) tabulated cross section values for collision energy from 1 keV/u to 630 keV/u. Schultz et al. (2023) improved the calculations from Schultz et al. (2016), which changed the cross section by up to ~3% near 1 keV/u. The two sources used in our fit agree within ~2% between 0.5 and 2 keV/u. We fit the formula to the data using the least-squares method. However, we weigh the Barnett (1990) data in the fitting procedure to provide uniform weight per logarithm of energy despite the much finer energy resolution in Schultz et al. (2023). The best-fit parameters are: $A_1$ = 4.8569, $A_2$ = 21.906, $A_3$ = 31.487, $A_4$ = 0.12018, $A_5$ = $4.1402\times10^{-6}$, $A_6$ = 3.7524, $A_7$ = $8.8476\times10^{-12}$, $A_8$ = 6.1091.

Figure A1 compares the cross sections from the discussed sources and the new fit formula. The new fit reproduces the Schultz et al. (2023) cross section within ±5% for collision energies from ~1 eV/u to 2 keV/u. The quantum mechanical oscillations of the cross sections below 1 eV/u cannot be reproduced. The fit reproduces the Barnett (1990) tabular data within ±5% for energy above 1 keV/u. The cross section reproduced using the Chebyshev polynomial coefficients is less accurate than the new formula. The Lindsay & Stebbings (2005) cross section is up to ~25% higher for the collision energy of ~20 keV/u compared to the new fit and the Barnett (1990) recommended value. The Janev & Smith (1993) cross section closely follows the Barnett (1990) values except for the highest collision energies, which were constrained by other sources. While the Swaczyna et al. (2019b) formula can be used to estimate the cross section for collision energy up to ~7 keV/u, it overestimates the cross section by up to ~30% compared to the Barnett (1990) tabular data at ~20 keV/u.

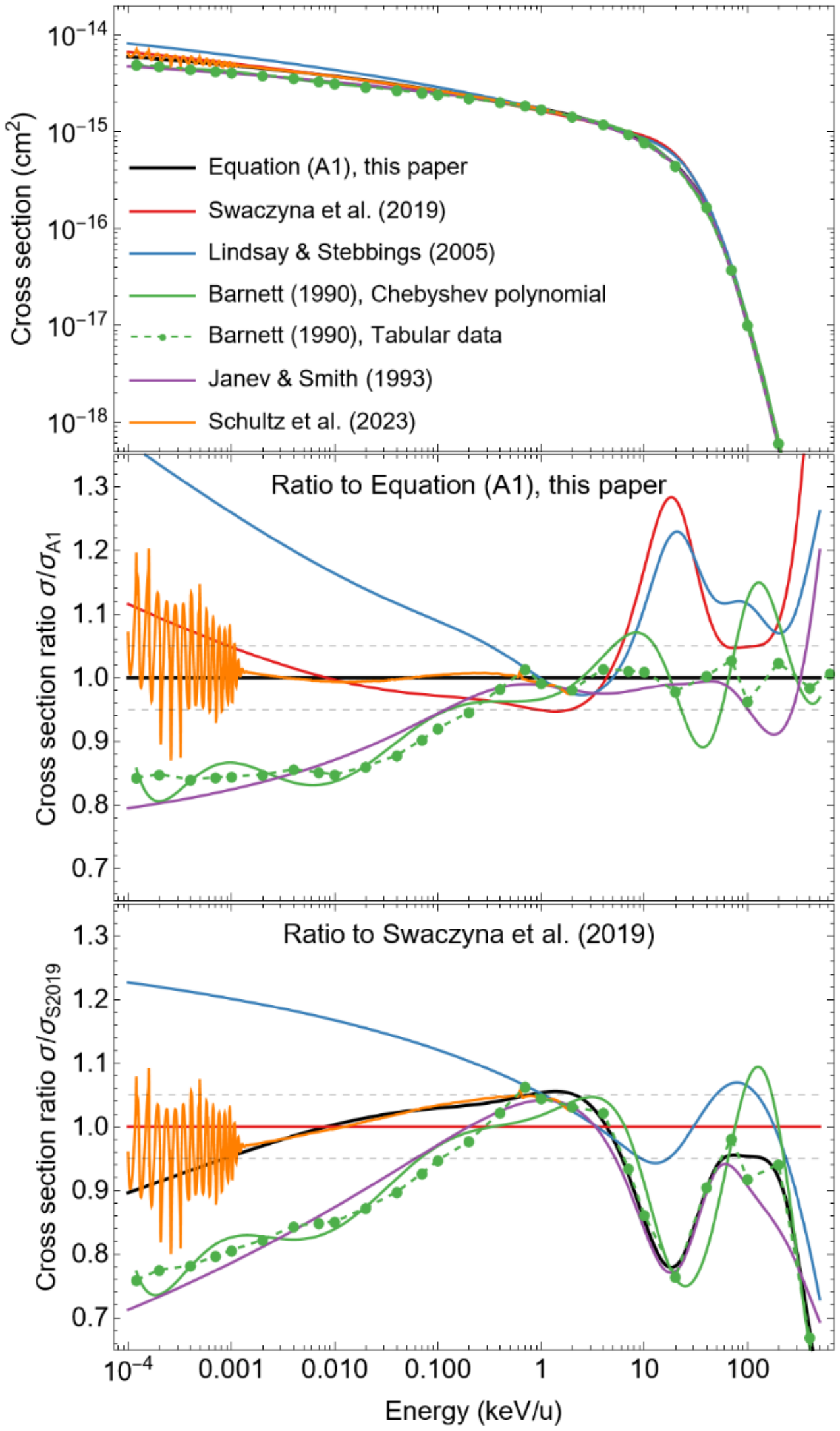

**Figure A1.** Comparison of the proton – hydrogen atom charge exchange cross sections considered in this study. The top panel compares the absolute cross sections. The middle and bottom panels show ratios to the recommended formula given in Equation (A1) and from Swaczyna et al. (2019b), respectively.

We recommend using the formula in Equation (A1) to calculate the proton – hydrogen atom charge exchange cross section over the energy range from 1 eV/u to 500 keV/u, covering the entire energy range observed by IMAP ENA detectors. We estimate the accuracy of the cross section to be 20% for collision energy below 1 keV/u based on discrepancies between the cross section sources. The accuracy above this range is adopted after Barnett (1990): 5% from 1 keV/u to 100 keV/u, and 20% above 100 keV/u.

## Appendix B. Cross section for $H^0 + H^0 \rightarrow H^+ + H^0 + e^-$

We adopt the same functional form as for the ionization in collisions with protons (Section 5.8) following Janev & Smith (1993, Section 2.2.1) to find an analytic approximation of the Cariatore & Schultz (2021) recommendation. The cross section σ in $cm^2$ is calculated from the collisional energy $E$ in keV/u as:

$$\sigma(E) = 10^{-16} B_1 \left[ \frac{\exp(-B_2/E)\ln(1 + B_3 E)}{E} + \frac{B_4 \exp(-B_5 E)}{E^{B_6} + B_7 E^{B_8}} \right]. \tag{B1}$$

We find the following parameters using the least-squares fitting of the logarithms of the cross sections: $B_1$ = 14.897, $B_2$ = 16.097, $B_3$ = 0.96601, $B_4$ = 5627.5, $B_5$ = 0.63515, $B_6$ = -6.5455, $B_7$ = $4.7441\times10^5$, $B_8$ = -1.8052. This formula reproduces the recommended values from Cariatore & Schultz (2021) with ±15% accuracy within the energy range from 30 eV/u to 10 MeV/u. The accuracy is ±5% for the collision energy above ~10 keV/u. Figure B1 compares the recommended value with the analytic formulae given in Equation (B1) and by Barnett (1990, reaction E-2). While there is a good agreement for high energies, a large discrepancy is seen below 10 keV/u.

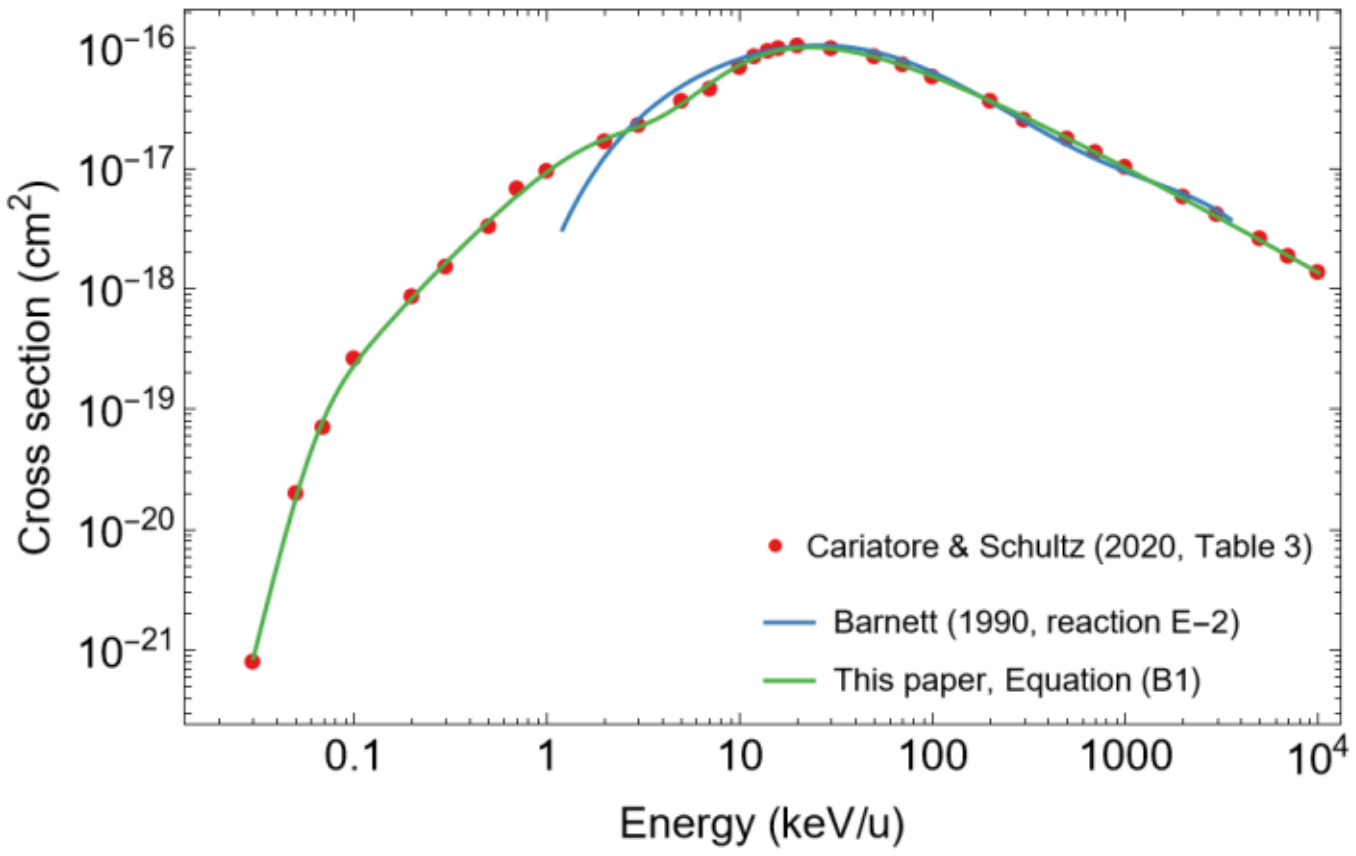

**Figure B1.** Comparison of the recommended cross section values from Cariatore & Schultz (2021) with the obtained fit formula and the cross section reproduced using the Chebyshev polynomial coefficients from Barnett (1990, reaction E-2).